# Tunable entanglement and strength in "granular metamaterials" based on staple-like particles: Experiments and discrete element models

Saeed Pezeshki [1], Youhan Sohn [1], Vivien Fouquet [1,2] and Francois Barthelat [1,*]

[1] Laboratory for Advanced Materials and Bioinspiration, Department of Mechanical Engineering, University of Colorado, 427 UCB, 1111 Engineering Dr, Boulder, CO 80309, USA

[2] Ecole Normale Supérieure Paris-Saclay, Université Paris-Saclay, 4 Avenue des Sciences, Gif-sur-Yvette

* Corresponding author: francois.barthelat@colorado.edu

**Abstract**

Entangled matter displays unusual and attractive properties and mechanisms: tensile strength, capabilities for assembly and disassembly, damage tolerance. While some of the attributes and mechanisms share some traits with traditional granular materials, fewer studies have focused on entanglement and strength and there are large gaps in our understanding of the mechanics of these materials. In this report we focus on the tensile properties and mechanics of bundles made of staple-like particles, and particularly on the effect of adjusting the angle between the legs and the crown in individual staples. Our experiments, combined with discrete element models, show competing mechanisms between entanglement strength and geometric engagement between particles, giving rise to an optimum crown-leg angle that maximizes strength. We also show that tensile forces are transmitted by a small fraction of the staples, which is organized in only 1-3 force chains. The formation and breakage of these chains is highly dynamic: as force chains break, they are replaced by fresh ones which were previously mechanically invisible. Entangled matter as "granular metamaterials" offer interesting perspectives in terms of materials design, and a vast design space for individual particles. Since their properties can be tuned with the shape of the staple, we interpret these entangled materials are "granular metamaterials" with unusual combination of properties: simultaneous strength and toughness, controlled assembly and disassembly, re-conformability, recyclability.

**I. Introduction**

Granular materials can be defined as a large collection of discrete, macroscopic particles which are in general non-cohesive, and they are ubiquitous in everyday life and in industry (food, pharmaceutics, powder metallurgy, geology, mining…). Individual grains or particles are typically large enough to neglect the effects of thermal agitation, and their main interaction mechanism is by frictional contact forces (Papadopoulos et al., 2018). Granular materials are seemingly simple systems, but they in fact display a wide range of nonlinear complex mechanical responses and behaviors including transition from "liquid" to "solid" states through jamming, or rich statistical mechanics (Behringer and Chakraborty, 2019; Jaeger et al., 1996). Typical granular materials are based on spherical grains, which is sub-optimal in terms of mechanical performance because of poor packing and localized "force lines" that occupy only a small volume in the material (Nicolas et al., 2000). The strength of granular material can be increased by tailoring the shape of the individual grains (Tsai and Gollub, 2004), or by assembling the grains into highly ordered "granular crystals" (Karuriya and Barthelat, 2023; Onoda and Liniger, 1990). Still, these granular materials can only sustain combinations of shear and compression. To generate tensile strength, adhesives may be added at the interface between the grains (Gans et al., 2020; Karuriya and Barthelat, 2024) or the shape of the grains may be enriched to generate geometrical interlocking. For example, long rods and star-like particles with slander branches create "weak" interlocking and non-negligible tensile strength (Aponte et al., 2024; Barés et al., 2017; Philipse, 1996). On the other extreme, in "stiff" fabrics akin to chainmail, the particle have loop topologies and they interweave with their neighbors, creating high tensile strength (Wang et al., 2021). An intermediate type of tensile granular matter is based on particles with hook-like or barb-like features that can entangle with other particles in a way that is tunable and reversible. For example, staple-like U shaped particles entangle to create structures which are free standing with high angle of repose (Gravish et al., 2012), structures with relatively high tensile strength and large deformability (Franklin, 2014; Gravish et al., 2012; Marschall et al., 2015) , and free standing granular "beams" with flexural strength (Murphy et al., 2016). Interestingly, the entanglement response and the strength of these "granular metamaterials" can be tuned with the geometry of the individual particles. For example longer "legs" in staple-like U-shape particles promote the "strength" of entanglement, but they also decrease the packing factor, so there is a subtle optimum in strength resulting from the competition between these two effects (Gravish et al., 2012). The legs can also be "twisted" about the backbone or "crown" of the staples to produce "Z-shaped particles", with an optimum twist angle of 90º (Murphy et al., 2016). When subjected to mechanical stresses, recent simulations have revealed the formation of tensile force lines within bundles of S-shaped and C-shaped particles, although these tensile force chains are sparser and less stable compared to compressive force chains (Karapiperis et al., 2022). Interestingly, nature also abounds in examples of entangled matter, such as bird nests (Weiner et al., 2020), fire ant rafts (Mlot et al., 2011; Wagner et al., 2021) and other morphological entanglement in living systems (Day et al., 2024). This existing research shows how control and tuning of particle geometry can lead to unusual and attractive mechanical properties. To this day, however, there is still a lack of fundamental understanding of the mechanisms underlying entanglement, strength and disentanglement and mechanics at regimes of large deformations, which makes it difficult to design and optimize particles to achieve specific mechanical performances. This report focuses on staple like particles where we manipulate the angle between the legs and the crown. We show, using experiments and discrete element models, that this simple geometrical change has profound implications on mechanical properties. We also use this model "granular metamaterial" as a platform to explore deformation mechanisms at regimes of large tensile deformation.

## II. Experiments

For this study we used standard, U-shaped steel office staples (Swingline, IL), because of their availability, and the ease with which their shape can be manipulated. We prepared bundles of staples from standard "sticks" of staples by immersion in an acetone bath, which in a few seconds dissolved the weak adhesive holding the staples together (Fig. 1a). Fig. 1b shows a schematic of the individual staple, with dimensions. The main parameter we altered was the angle between the legs and the crown $\theta$ (the "crown-leg" angle). To change the crown leg-angle from the standard $\theta=90°$ in an efficient way, we 3D printed a series of tools with various geometries to either close (Fig. 1c) or open (Fig, 1d) the crown leg angle of a whole stick of staples. Once the legs were bent to the desired angle, we separated the individual staples with acetone. Using this protocol, we prepared batches of staples with different crown-leg angles ranging from $\theta=20°$ to $\theta=120°$ (Fig. 1).

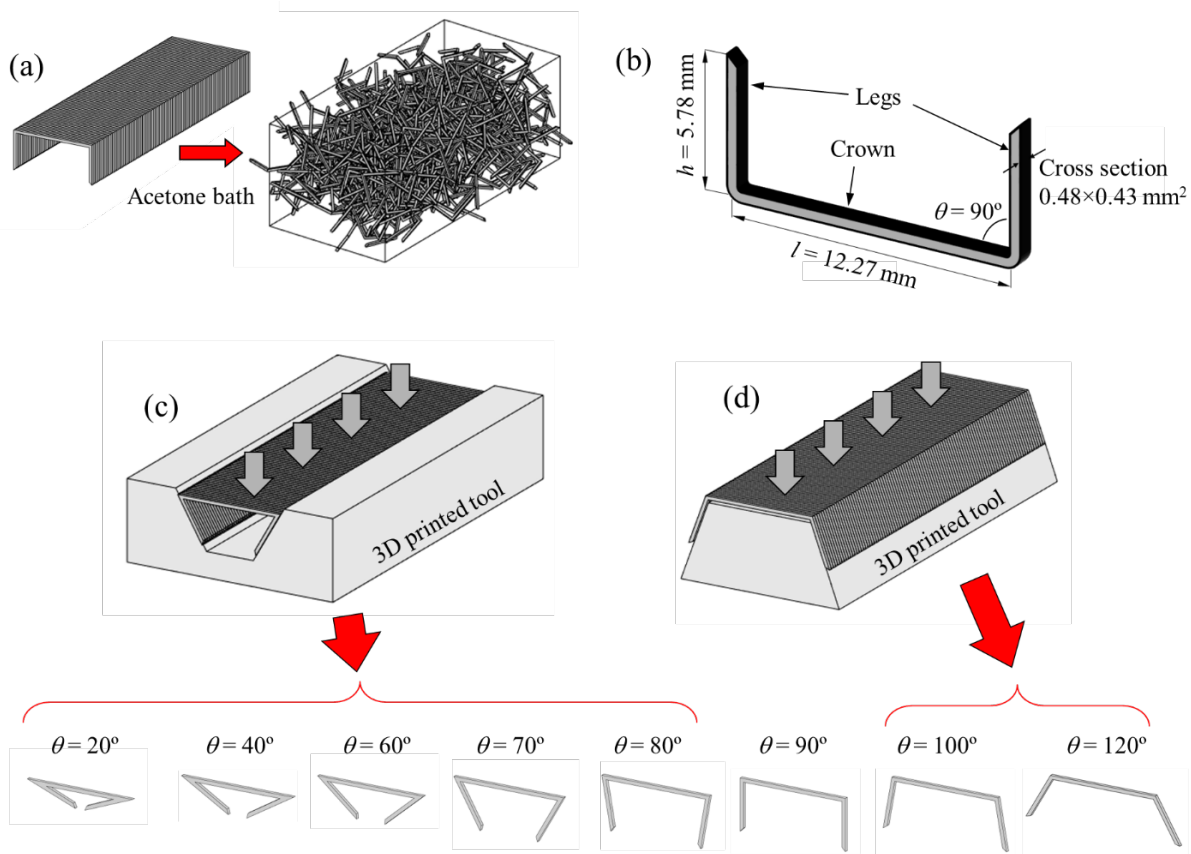

Fig. 1. Sample preparation: (a) An acetone bath disassembles a "stick" of staples into individual staples; (b) Dimensions of an individual staple with a crown-leg angle $\theta$=90°. The crown-leg angle $\theta$ can be either (c) closed or (d) open using custom 3D printed tools.

Intuitively, changing the crown-angle should have an effect on the entanglement strength and on the tensile strength of a bundle made of these staples, since at a fundamental level the entanglement strength is generated by pairs of individual staples latching on one another through a hook-like mechanism. To quantify this entanglement strength, two staples were clamped in a miniature tensile machine (ADMET eXpert 4000 Micro Tester), and a third staple was placed in a fully latched configuration between the two clamped staples. This formed a configuration where the crowns of the three staples were the closest to the line of action of the pull force. The two outer staples were then slowly moved apart at a quasi-static rate of 10 mm /min while the force was recorded with a 500 N load cell. Fig. 2a shows typical force-extension curves for staples with $\theta=40°$, $\theta=60°$, and $\theta=90°$. The main deformation mechanism of these staples in tension was the progressive elastic, then plastic bending of the staples at the elbows between the crown and the legs until a peak force was reached, which was followed by a more or less abrupt drop corresponding to the center staple sliding or shifting position. The test often concluded with the center staple being "ejected" from the chain from instability and elastic snapback.

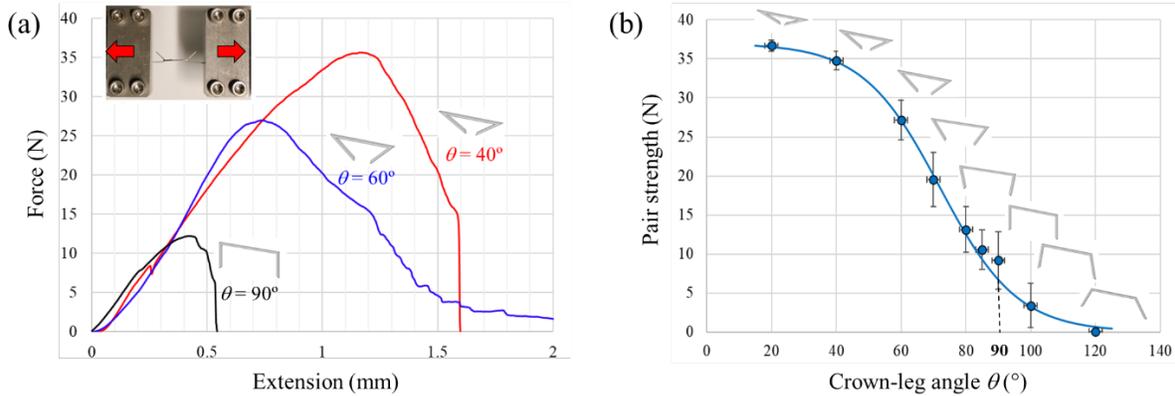

Fig. 2. (a) Typical force- extension curve for tensile tests on staple pairs and for three different crown-leg angles; (b) Pair strength as function of the crown-leg angle.

As expected, reducing the angle $\theta$ resulted in higher peak forces ("pair strength") and also in higher deformability, because more deformation was needed to bring the staples in an unstable configuration. The results for maximum force or "pair strength" are shown on Fig. 2b (10 staple chains were tested for each angle). The average pair strength for three staples with 90° crown-leg angle (standard office staple) is about 9 N. Closing the crown-leg angle from $\theta = 90°$ significantly increased the strength, up to about 37 N for $\theta =20°$ staples. On the other hand, increasing the crown-leg angle from $\theta = 90°$ led to a rapidly vanishing pair strength.

The next series of experiments was on bundles of these staples. For each crown-leg angle, 2000 staples were prepared and poured into a 100×42×35 mm³ acrylic box. The box was then placed on a vertical vibration stand (Eisco Labs Vibration Generator), which was vibrated sinusoidally for 30 seconds using a sine wave signal generator (Gravish et al., 2012). The intensity of this combination of vibration under gravity was quantified by the nondimensional number $\Gamma=A\omega^2/g$ (Li et al., 2020) where $A$ is the amplitude, $\omega$ is the angular frequency and $g$ is the gravity constant. In this study we used $\Gamma=2.0$, which is a typical value for granular materials (Gravish et al., 2012). However, we found that these vibrations had negligible effects on the packing factor of the staples (solid volume / total volume), which we estimated at around 0.1 (which compares well with previous studies on similar particles (Gravish et al., 2012). The limited effects of vibration in our case can be explained by existing local entanglement in the bundles and by the mechanical confinements of the container, which together severely restrict the mobility and possible rearrangements of individual staples within the bundle. To perform tensile tests on bundles of this staples we transferred the pile onto a horizontal Teflon substrate and between the crosshead of a horizontal tensile machine (ADMET eXpert 4000 Micro Tester). This bundle of staples could not be clamped like traditional tensile test samples so instead, we fabricated a pair of custom "claws" made of six steel nails. These claws were driven through the end of the bundle and then attached to the loading machine (Fig. 3a) to "grab" the ends of the bundle and apply tensile deformations. The initial distance between the claws was 65 mm for all samples and their cross section was approximately 42 mm × 19 mm. Ten samples were tested for each crown-leg angle explored. All tensile tests were conducted in a quasi-static pulling rate of 10 mm/min.

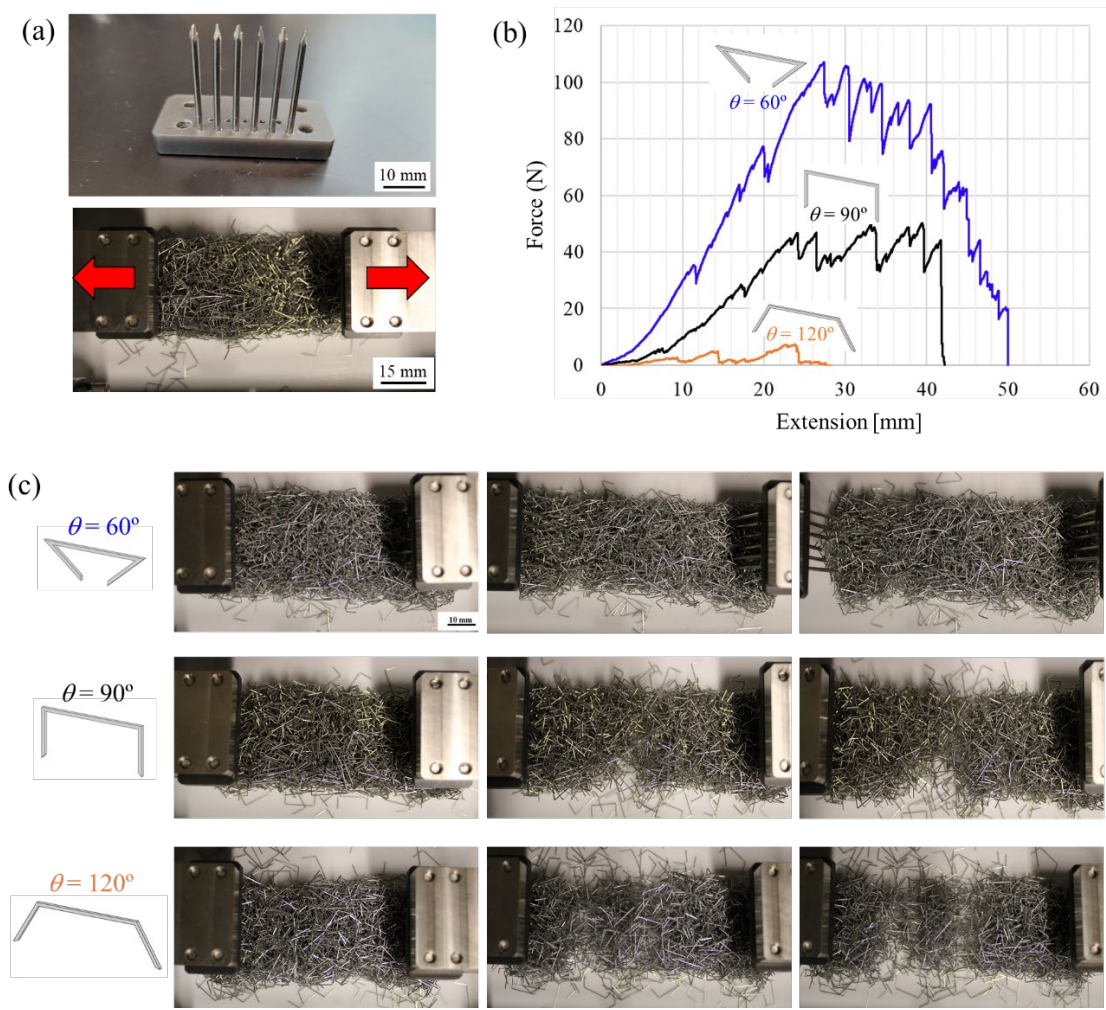

Fig. 3. Tensile tests on bundles of staples: (a) Custom made "claw" and tensile setup; (b) typical force-extension curves and (c) Snapshots of typical tensile tests for staples with three different crown-leg angles.

Fig. 3b shows typical tensile force-extension curves for bundles made of staples with $\theta = 60°$, 90°, and 120° crown-leg angles. In all samples the response started with a roughly linear force extension curve, followed by non-linear deformation and a highly jagged response consistent with previous tensile tests on staple-like particles (Franklin, 2014). Images acquired during these tensile tests showed a relatively homogeneous deformation (Fig 3c) and also staples which were regularly 'ejected" from the bundle during the tensile test, suggesting that the breakage and snapback of entangled staples during deformation. For $\theta = 90°$ and $\theta = 120°$ bundles, deformations eventually localized into one or two regions, and complete tensile failure occurred at strains in the 50% to 75% range. The $\theta = 60°$ bundles were more stable, producing relatively homogeneous deformations. The deformability of these bundles (and of bundles with $\theta < 90°$) was also much larger and could not be measured with our setup: At the maximum extension allowed by our loading machine ($u_{max}$=50 mm) these bundles had not entirely failed. Fig. 4 shows the bundle strength and as function of crown-leg angle. The expectation was for the strength to increase as $\theta$ was decreased, in a way consistent with the results for the staple pair. Instead, the strength showed a peak strength for staples in the $\theta = 40°$ to 60° range, with a strength up to six times greater than bundles made of staples with $\theta = 20°$ and $\theta = 120°$ staples.

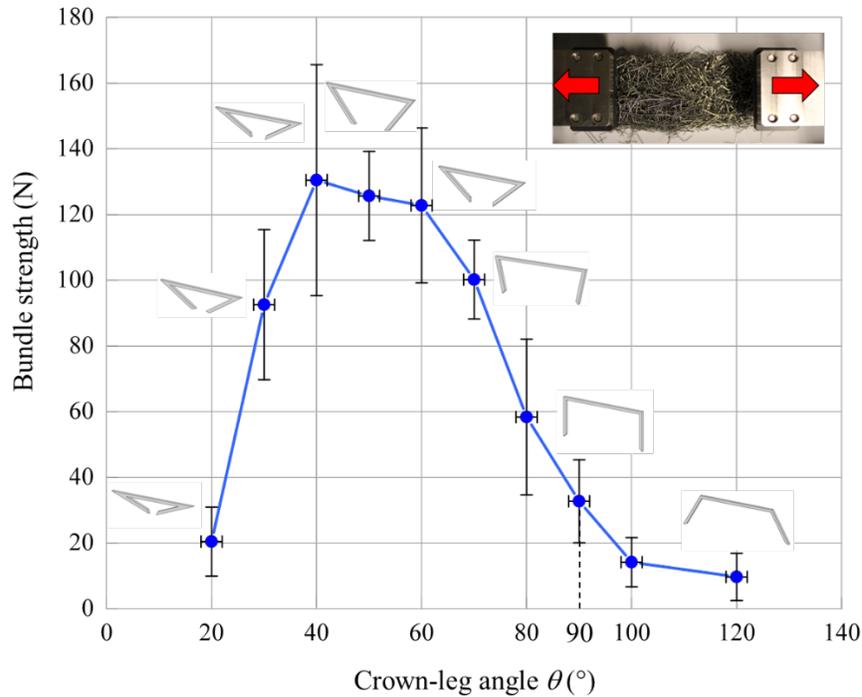

Fig. 4. Bundle tensile strength as a function of crown-leg angle for individual staples

We now summarize the main observations from the experiments, and we formulate some hypotheses on the mechanisms that govern the observed properties and trends. The entanglement strength of a pair of staples ("pair strength") increases significantly and monotonically as the crown-leg angle $\theta$ is decreased. However the strength of bundles of these staples shows a peak at $\theta$ =40° to 60°. Bundles made with staples with $\theta$ >90° are understandingly weaker, because these staple produce very weak entanglements (Fig. 2b). To explain the weak response of bundles with $\theta$ < 40°, we hypothesized that these staples do not engage and entangle as well as other staples with more "open" geometries. We also note that the strength of a bundle of 2000 staples is only 4-6 times greater than the pair strength of the same staples. We hypothesize that this relatively low strength is due to imperfect load transfers: Most staples in the bundles are not aligned with the direction of pull, and the tensile forces may be transferred only through a few force chains within the bundle. Furthermore, the jagged characteristic of the force- extension curves for the bundle suggests that these chains may break while other new chains may form, so that the process of force transfer and the structure of the force chains is highly dynamic and evolves with deformation. A few chains would then explain the relatively low strength of the bundle compared to the staple pair, but the dynamic process of breakage and re-formation can lead to large deformations, and large overall toughness for the bundle. For the second part of this study, we used discrete element models to explore these mechanisms and hypotheses.

### III. Discrete element models (DEM)

The main objective of the models presented here was to develop some understanding of the mechanics of entanglement, disentanglement and force transfer in the staple bundles. The discrete element methods, initially developed for traditional granular materials (Burman et al., 1980; Cundall and Strack, 1979), is particularly well suited for this discrete system of staples. We used the granular mechanics package of LAMMPS (Thompson et al., 2022), with a modeling approach where individual staples were discretized with spheres (Garcia et al., 2009; Karuriya and Barthelat, 2024; Kruggel-Emden et al., 2008; Li et al., 2015; Murphy et al., 2016) (Fig. 5). The overall geometry was identical to the staples tested above, with a crown length $l$=12.27 mm, a leg length $w$=5.78 mm and a tunable crown-leg angle $\theta$. The backbone of the staple was discretized with spheres of diameter $d$=0.45 mm to approximate the 0.48×0.43 mm$^2$ cross section of the staples, with a spacing $s$ along the backbone.

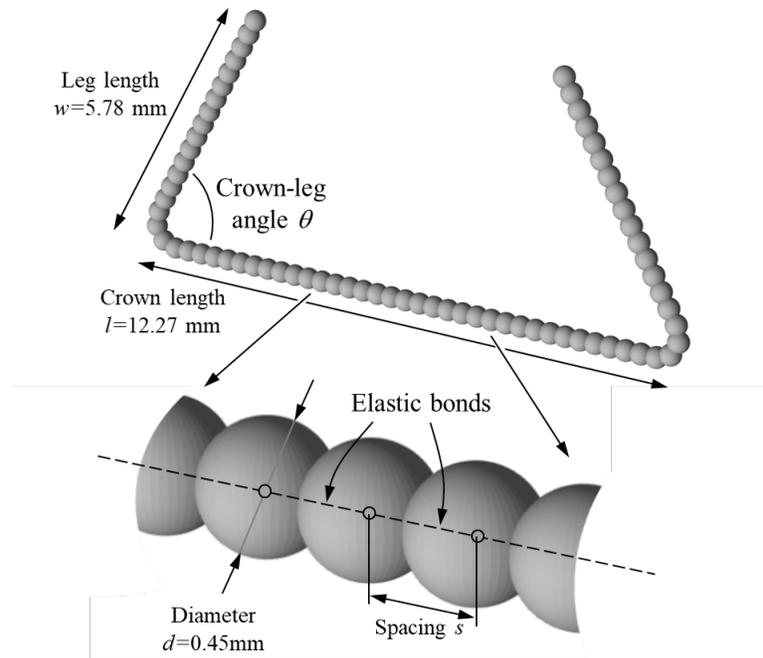

Fig. 5. DEM model of an individual staple based on discrete spheres joined with elastic bonds.

The experiments on staple chains above show that the deformation of the individual staples is an important disentanglement mechanism. To capture this deformation, the spheres were joined by elastic elements along the backbone, using the bonded particle models (BPM) in LAMMPS. The stiffnesses of the elements were computed from the modulus of the staple backbone, the spacing of the spheres $s$ and the properties of the cross section. These elements could deform in bending (stiffness $k_b = EI/s$ with $I$ =second moment of area), but also axially (stiffness $k_r = AE/s$ with $A$

=cross sectional area ), in torsion (stiffness $k_t = JG/s$ with $G$=shear modulus and $J$=polar moment of area) and in shear (stiffness $k_s = AG/s$ ). The strength of these bonds was assumed to be infinite. Spheres from different staples interacted by Hertzian contact with a linear history frictional model, and frictional forces between the particles that followed Brilliantov et al. (Brilliantov et al., 1996), Silbert et al. (Leonardo E. Silbert Deniz Ertaş and Plimpton, 2001), and Zhang et al (Zhang and Makse, 2005). The normal contact stiffness was estimated using $K_n$ =1×10$^5$ N/m, and the friction coefficient was set to $\mu$=0.3 (the static and dynamic friction coefficients were assumed to be identical). In our simulations, the interpenetration distance at the contact was negligible compared to the deformations of the staples, which occurred mainly in bending. The DEM method uses a Verlet time integration method, so we verified that the rate of deformation used in the simulations corresponded to "quasi-static" conditions where the results did not depend on the mass of the staples. An important numerical parameter is the sphere spacing *s* (Fig 5). In models with 2000 staples, decreasing *s* significantly increased the computational cost, but only with small (<10%) differences in predicted strength. We found that a spacing of *s=d* offered a good compromise between accuracy and computational cost.

*Pair Strength:* The first set of models presented here captures the three staple chains in tension and the "pair strength". Three staples were positioned identically to the experiment described above, and the two outer staples were separated at a constant rate. Gravity was turned off for these simulations.

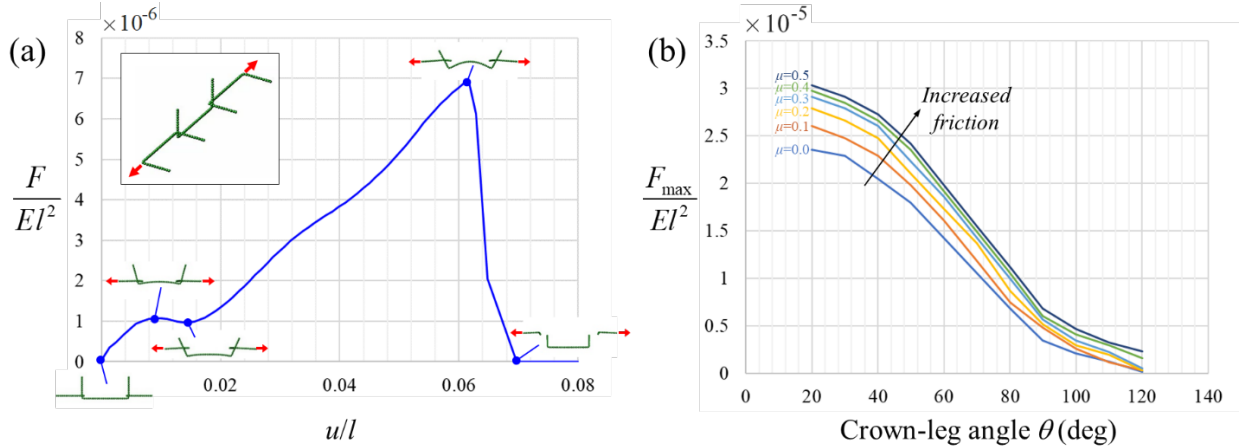

Fig. 6. Results from the DEM model of the three staples in tension: (a) Typical force-extension curve and snapshots of the model (shown for $\theta=90°$); (b) Pair strength as function of crown-leg angle, plotted for different friction coefficients $\mu$.

Fig. 6a shows a typical force-extension curve from this model. The force was normalized by $El^2$ where $E$ is the Young's modulus of the staples and $l$ is the length of the crown. The extension $u$ was normalized by $l$. The overall shape of the curve and the failure mode are identical to the experiments: A progressive rise of the force with extension as the legs bend mostly near the elbow between crown and legs, followed by a sudden snap of the central staple once the frictional resistance is overcome. Fig. 6b shows the normalized maximum force and the crown-leg angle for different friction coefficients. The results show that the staple pair strength follows the same trend as the experiment: the strength increases sharply as the initial crown-leg angle is reduced. Fig. 6 also shows that as expected, $F_{max}$ is higher for higher coefficients of friction ($\mu$). Frictional effects are however modest compared to the effects of geometrical interlocking. Even without friction ($\mu=0$), the staples produced a substantial tensile strength.

*Bundle strength:* We next present DEM models and results for the tensile deformation of bundles of 2000 staples. Staples were first randomly poured into a 100×42×35 mm³ box (Fig. 7a). The walls of the box and the staples interacted with a Hertz contact model with parameters identical to the ones used in the staple-staple interactions. Pouring, and the rest of the simulation, was also subjected to vertical gravity, which drove the staples into the box during the pouring step and kept the staples against the bottom surface of the box during the rest of simulations. Gravity was an important parameter in these simulations since it governed the "bouncing" response of the staples and some of the deformation of individual staples in the bundle. For the simulations we set the mass and the gravity so that $EI^2/mg = 10^8$, which produced the adequate response during the pouring process: Higher values for this ratio produced models with excessive bouncing of staples during the pouring process, and lower values produced models with excessive deformations, from gravity only, in the staples already deposited. This pouring process also produces a volume fraction of staples around 0.1, which is close to the experiments. Pulling on the bundle involved clamping of the ends of the bundle, and in the DEM models we adopted an approach which duplicated the experiment: once all 2000 staples were settled in the box we created six vertical rods within each end of the bundle, which interacted with the staples following the same law as the staple-wall interactions. The sudden incursion of the rods resulted in localized disturbances in the bundle, which we let relax in the simulations. Following this step, we removed the side wall of the box, while keeping the bottom floor and maintaining gravity. We then moved the sets of rods apart at a slow rate, so as to stretch the bundle quasi-statically. Fig. 7b shows the force-extension curve from these simulations, which displays all the characteristics of the experimental curve: Initial loading followed with a jagged nonlinear region, with numerous sudden unloading and slower reloading events. Snapshots of the stretched bundle (Fig. 7c) also capture the deformation and failure mode of the experiment: Homogenous deformations, followed by localization and failure.

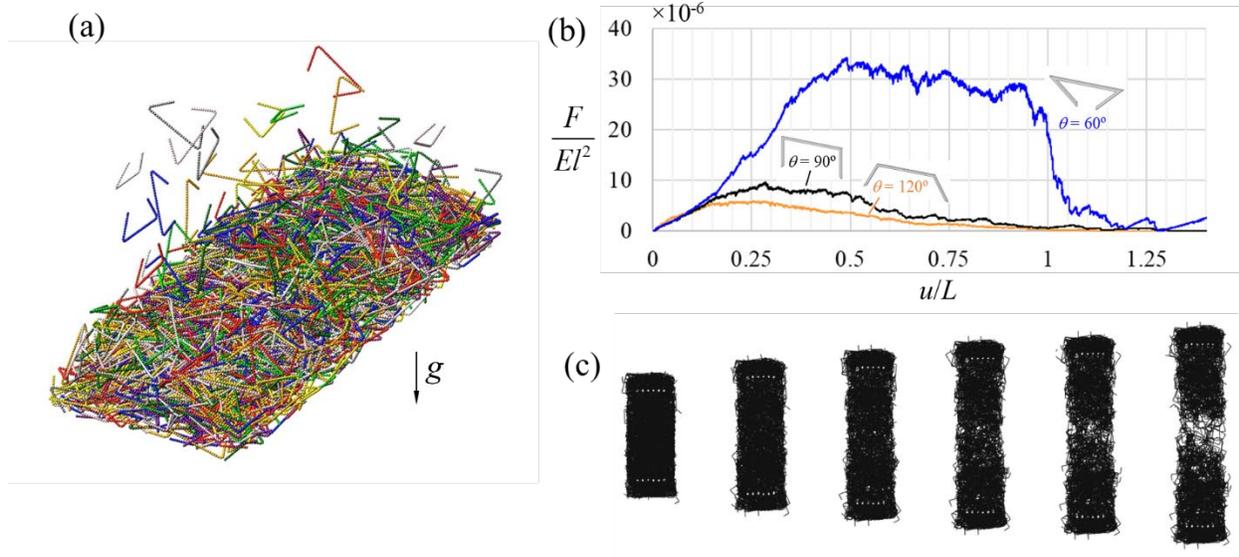

Fig. 7. DEM results for a bundle of 2000 staples: (a) Snapshot of the "pouring" step; (b) Normalized force-extension curve for bundles of $\theta=60°$, $\theta=90°$ and $\theta=120°$ staples; (c) Snapshots taken at different levels of deformation for the bundle with $\theta=90°$ staples.

The model also captures the strong dependance of the mechanical properties of the bundle on the crown-leg angle $\theta$. Fig. 8a shows the normalized maximum force in the bundle as function of crown-leg angle $\theta$. The results are consistent with experiments, with an optimum of strength at crown-leg angle $\theta=40°$, and much lower tensile strengths for $\theta=20°$ and $\theta=120°$. We also report, on Fig. 8b, the predicted strain at failure for the bundle. The trend is consistent with the experiment, with a higher deformability at the angle $\theta$ is reduced. An optimum strain at failure of 4.5 is obtained for $\theta=30°$, after which the strain at failure drops for $\theta=20°$. The strongest designs are therefore the more deformable and tough, which is an unusual and intriguing feature: Strength and toughness are usually mutually exclusive in engineering materials (Ritchie, 2011). We now analyze in more details the mechanisms of entanglement and force lines within the bundle.

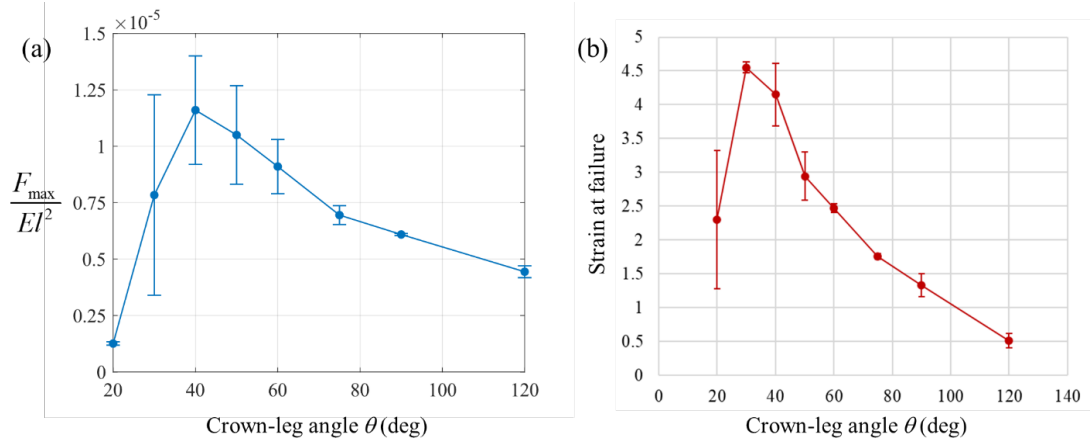

Fig. 8. DEM results for a bundle of staples: (a) Normalized maximum force and (b) strain at failure, both plotted as function of crown-leg angle $\theta$.

*Entanglement:* We used the DEM results to explore the dynamic of entanglement in the bundle. To this end, we used a simple entanglement criterion between pairs of staples. We defined "nets" for each staple, which are flat area partially enclosed by the crown and legs of the staples which can engage "catch" other staples in the bundle (Fig. 9a).

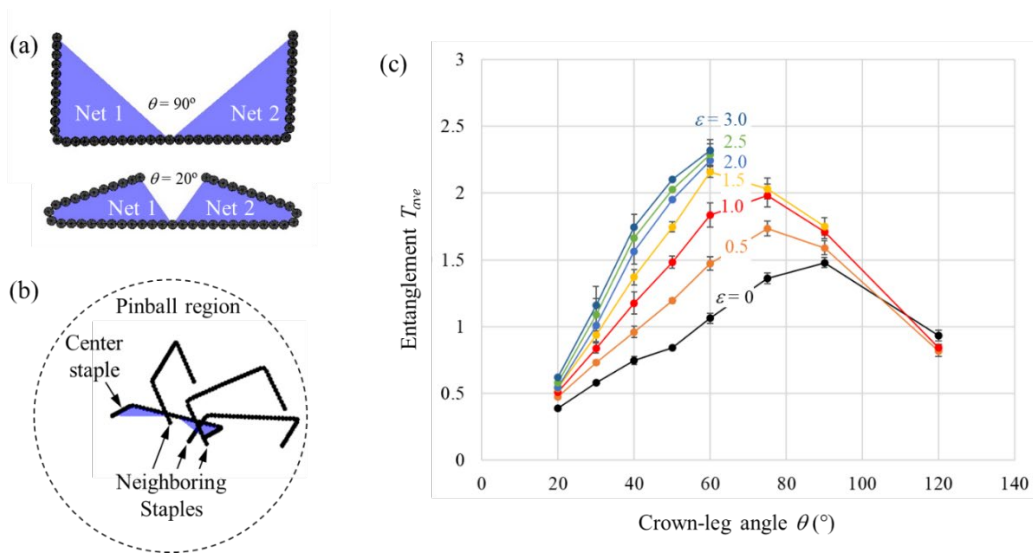

Fig. 9. Entanglement criterion: (a) Plane regions called "nets" are defined for each staple. These regions can "catch" neighboring staples for entanglement; (b) For each staple in the bundles, the number of neighboring staples crossing the nets are counted; (c) Average entanglement as function of crown-leg angle, plotted for different levels of tensile deformation.

For each staple in the bundle, we considered all neighboring staples within a pinball region of radius equal to two times the length of the crown, and centered on that staple of interest ("center staple", Fig. 9b). We then counted, using a simple three-dimensional intersection algorithm, the number $T$ of these neighboring staples that intersected any of the two nets of the center staple (Fig. 9b). For counting we enforced a "reciprocal" entanglement condition, only counting those neighboring staples whose nets also intersect the nets of the center staples. For a given conformation of the staple bundle, we finally computed $T_{ave}$, the average number of staples entangled with each staple in the bundle. Fig. 9c shows $T_{ave}$ as function of crown-leg angle, plotted for different levels of deformation in the bundle ($\varepsilon = 0$ to 3.0). The entanglement is generally the poorest for $\theta=120°$ because the net areas are small, and for $\theta=20°$ because the staples are too "closed" geometrically. This observation confirms one of our hypothesis and prediction: While the $\theta=20°$ staples have the strongest entanglement when they do entangle, their geometry is too "closed", so that they do not engage well with neighboring staples. This effect results in the relatively low bundle strength for $\theta=20°$ (Fig. 8a). It is therefore the combination of entanglement strength and geometrical entanglement in the bundle gives rise to the optimum bundle strength at around $\theta=40°$. Interestingly, Fig. 9c also shows that increasing strains in the bundle increases average entanglement, this effect being the most pronounced for staples at $\theta=60°$. Pulling on the bundle induces relative motion between neighboring staples at the local level, giving them additional opportunities to engage and entangle. We note that geometrical entanglement between staples does not mean that appreciable forces will be transmitted between these staples, which motivates the next and final section.

*Force lines:* We finally used the DEM results to explore how tensile forces are transmitted within the bundle. We first computed the average tensile force $F_i$ carried by each individual staple $i$ in the bundle. A typical cumulative distribution of these forces, normalized by the staple forces average over the entire bundle ($F_{ave}$), is shown in Fig. 10a for $\theta=90°$ and $\varepsilon=0.61$, which is representative of other staples. The results show that most of the staples (~65%) carry less than the average force in the bundle. We fitted this cumulative distribution can be fitted with a Burr type XII distribution with parameters $(\alpha,c,k)$ (Rodriguez, 1977) (Fig. 10a):

$$P(x)=1-\frac{1}{\left(1+(x/\alpha)^c\right)^k} \qquad (1)$$

Using these parameters we plotted the probability density function (pdf), together with a histogram of forces, on Fig. 10c. The pdf for the Burr distribution is:

$$p(x)=\frac{(kc/\alpha)(x/\alpha)^{c-1}}{\left(1+(x/\alpha)^c\right)^{k+1}} \qquad (2)$$

These distributions are wide with a large positive skew and relatively long tail, suggesting that only a small subset of staples carry high forces.

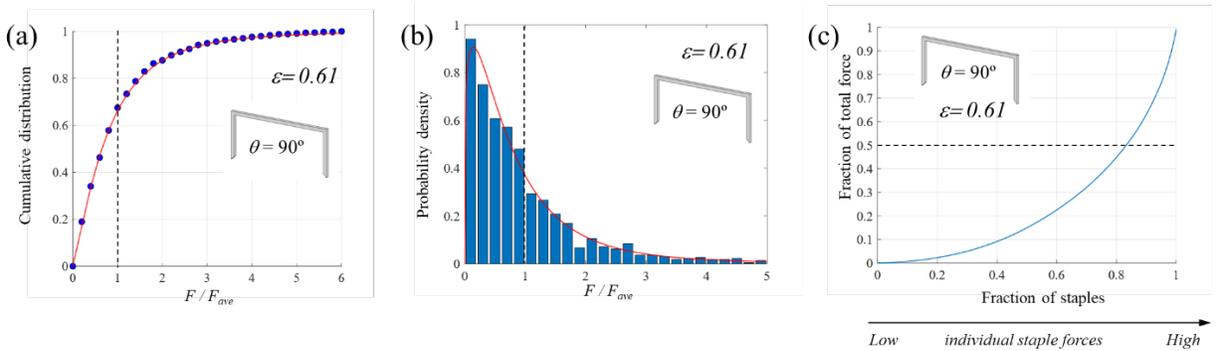

Fig. 10. (a) Cumulative distribution and Burr distribution fit for normalized forces in individual staples (case $\theta=90°$, $\varepsilon=0.61$); (b) Histogram with Burr probability density function; (c) Fraction of total force as function of fraction of staples for the case $\theta=90°$ and $\varepsilon=0.61$.

This observation brings the question of which fraction of the total tensile force is carried by this small subset of highly loaded staples. We ranked the staples according to how much force they carry, and then we computed the fraction of the total force that a cumulative fraction of these staples carried. The results, showed on Fig. 10c for 90° staples, show that at tensile strain of $\varepsilon =$ 0.61, about 85% of the lowest load carrying staples carry half of the total force in the bundle. In other words, only the 15% highest load carrying staples carry half of the total force. At larger deformation, this effect is even more pronounced, and these values and trends are similar for bundles with staples $\theta$<90°. These statistics, taken over the bundle and at different levels of deformations, strongly suggest that only a few force lines carry the applied stress, as observed in previous simulations (Karapiperis et al., 2022). We now focus on the staples that carry the most forces in the bundle, and more specifically to the subset of staples that carry, as a group, one third of the total applied force on the bundle. Fig. 11 shows the results for $\theta$=60° staples, which are representative of the mechanics of force transmission in the strongest staple designs. We first note that high force-bearing staples organize in only one to three force chains, which explains the relative low strength of a bundle of 2000 staples compared to the pair strength. We also observe that force lines appeared and disappeared dynamically as deformation is increased. On Fig. 11 the staples were color-coded accordingly: the staples that just joined the group of highly loaded staples (newly born) are in green, and the staple is about to be leave that group (in the next increment of deformation) are in black. Staples that have a persistent presence before, during and the current deformation step are shown in blue (persistent staples). The formation and breakage of force lines is indeed a highly dynamic process. Early in the deformation, high forces staples are unloaded only shortly after they appear. A few chains then form, persist for a small range of deformation, and then break, which is associated with a sudden drop in the tensile force. However, these "broken" chains are immediately replaced by new chains, so that force increases again with extension. This highlights a mechanism where as major force lines fail, new force lines take over from mechanically "hidden lengths" in the bundle. This mechanism produces large deformations, some toughness, and the sawtooth pattern on the force-extension curves. Similar mechanisms of "sacrificial bonds" and "hidden molecular lengths" are observed at the molecular scale in tough proteins found in nacre (Smith et al., 1999) and bone (Fantner et al., 2005).

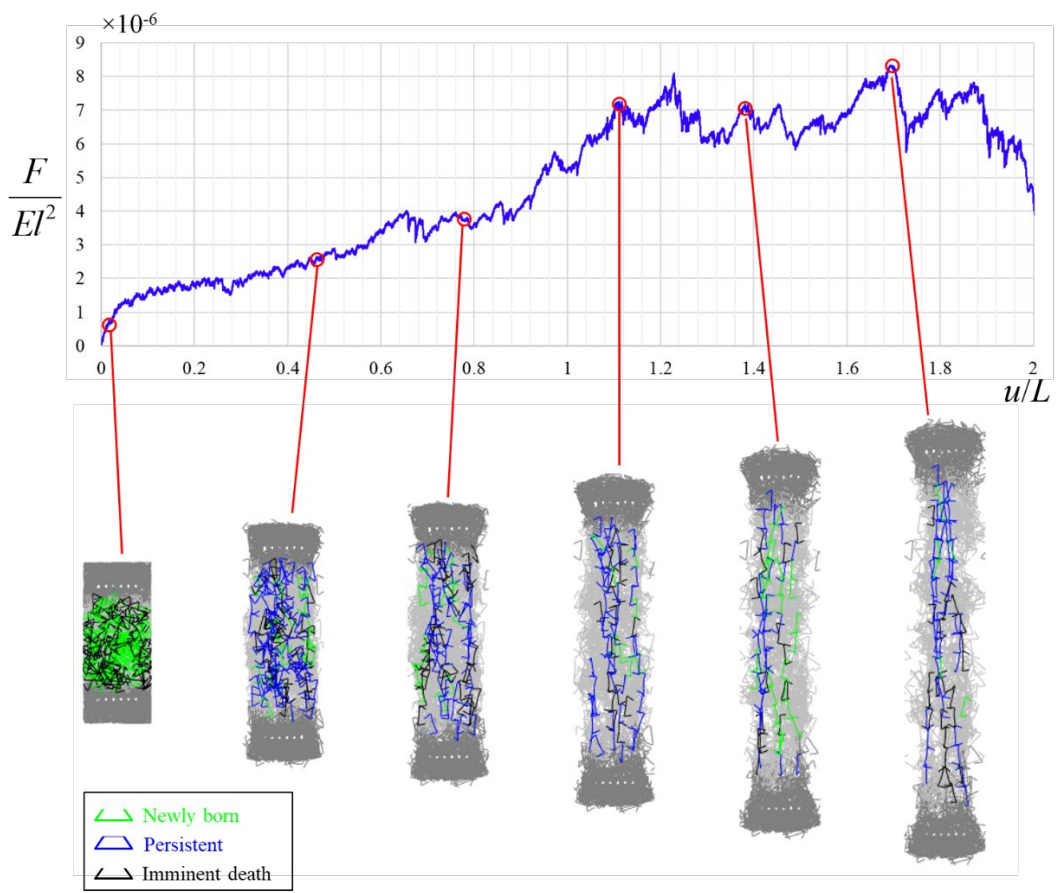

Fig. 11. Force lines in a stretched bundle of $\theta=60°$ staples. The colored group of staples carry 1/3 of the entire tensile force. These staples are color-coded according to whether they just joined that group (newly born, green), stay in that group (persistent, blue) or are about to leave that group (imminent death, black) in the next deformation increment.

## IV. Summary

Entangled matter displays unusual and attractive properties and mechanisms: tensile strength, assembly and disassembly. In this report we focus on staple-like particles, and on the effects of tuning the crown-leg angle. We show that the strength of interaction between pairs of staples, and the mechanical properties of a bundle of these staples, is highly sensitive to this crown leg-angle. The discrete element model presented here captures both staple-pair tensile experiments and tensile tests on bundles of 2000 particles. To reach that level of fidelity with the experiments, modeling individual staples as flexible structure was required (in contrast with previous models that assumed rigid particles (Karapiperis et al., 2022)). The experiments and models collectively point to the following main conclusions:

- Staple pair strength increases when the crown-leg angle is decreased.
- The amount of geometrical engagement and entangled between staples in the bundle is a strong function of the staple geometry. In particular, when the geometry of the staples is too "closed" ($\theta=20°$), entanglement with neighboring staples is limited.
- The tensile strength of the bundle is governed by the pair strength of the individual staples but also by their propensity to entanglement with their neighbors in the bundle. These competing effects give rise to an optimum design around $\theta=40°$ - $60°$ for the crown-leg angle.
- Average entanglement in the bundle increases with tensile deformations, because idle staples are "recruited" by the relative motion of staples as the bundle deforms.
- The tensile forces are transmitted by a small fraction of the staples, which are organized into only 1-3 force chains in the bundle (in consistence with previous studies (Karapiperis et al., 2022). The formation and breakage of these chains is a highly dynamic process: as force chains break, they are replaced by fresh ones which were previously mechanically invisible ("hidden lengths").

Since the mechanical properties of the bundle can be tuned with the shape of the staple, we interpret these entangled materials are "granular metamaterials" with unusual combination of properties. For example, we show that adjusting the crown-leg angle can result in an increase of both strength and deformability, which is unusual since these properties are mutually exclusive in traditional engineering materials (Ritchie, 2011). Entangled matter as "granular metamaterials" offer a rich landscape in terms of mechanics and vast perspectives in terms of design. The geometry of the individual particles may be enriched to maximize entanglement and strength. Other parameters such as pouring protocol, container size and shape, vibration type and intensity can also probably be optimized to maximize strength, deformability and toughness. Manipulating the history of the bundle, for example by pulling the bundle along a certain direction to pre-condition the material and promote the formation of force lines ("drawing") also offers interesting perspectives. The present study has only explored a small fraction of this vast design space and possible processing steps, and it likely that more complex three-dimensional designs for individual particle, and optimized preparation protocols, will lead to improved properties. The strength of these materials may not ever reach the strength of monolithic engineering materials, but some of their unique features (versatile construction, assembly, disassembly and recycling, damage tolerance, unusual combination of strength and toughness) will make them interesting for some applications.

**Acknowledgments**

This work was supported by the US National Science Foundation (Mechanics and Materials and Structures, Award No. 2033991). SP and YS were also partially supported by the Department of Mechanical Engineering at the University of Colorado. VF's research internship was supported by the Ecole Normale Superieure de Paris-Saclay. The DEM simulations were performed on the Alpine high performance computing resource at the University of Colorado Boulder. Alpine is jointly funded by the University of Colorado Boulder, the University of Colorado Anschutz, Colorado State University, and the National Science Foundation (Award 2201538).

**References**

Aponte, D., Estrada, N., Barés, J., Renouf, M., Azéma, E., 2024. Geometric cohesion in two-dimensional systems composed of star-shaped particles. Phys Rev E 109, 44908. https://doi.org/10.1103/PhysRevE.109.044908

Barés, J., Zhao, Y., Renouf, M., Dierichs, K., Behringer, R., 2017. Structure of hexapod 3D packings: Understanding the global stability from the local organization. EPJ Web Conf 140, 3–6. https://doi.org/10.1051/epjconf/201714006021

Behringer, R.P., Chakraborty, B., 2019. The physics of jamming for granular materials: a review. Rep Prog Phys 82, 12601. https://doi.org/10.1088/1361-6633/aadc3c

Brilliantov, N. V, Spahn, F., Hertzsch, J.-M., Pöschel, T., 1996. Model for collisions in granular gases. Phys. Rev. E 53, 5382–5392. https://doi.org/10.1103/PhysRevE.53.5382

Burman, B.C., Cundall, P.A., Strack, O.D.L., 1980. A discrete numerical model for granular assemblies. Geotechnique 30, 331–336. https://doi.org/10.1680/geot.1980.30.3.331

Cundall, P.A., Strack, O.D.L., 1979. A discrete numerical model for granular assemblies. Géotechnique 29, 47–65. https://doi.org/10.1680/geot.1979.29.1.47

Day, T.C., Zamani-Dahaj, S.A., Bozdag, G.O., Burnetti, A.J., Bingham, E.P., Conlin, P.L., Ratcliff, W.C., Yunker, P.J., 2024. Morphological Entanglement in Living Systems. Phys. Rev. X 14, 11008. https://doi.org/10.1103/PhysRevX.14.011008

Fantner, G.E., Hassenkam, T., Kindt, J.H., Weaver, J.C., Birkedal, H., Pechenik, L., Cutroni, J.A., Cidade, G.A.G., Stucky, G.D., Morse, D.E., Hansma, P.K., 2005. Sacrificial bonds and hidden length dissipate energy as mineralized fibrils separate during bone fracture. Nat Mater 4, 612–616. https://doi.org/10.1038/nmat1428

Franklin, S. V., 2014. Extensional rheology of entangled granular materials. EPL 106. https://doi.org/10.1209/0295-5075/106/58004

Gans, A., Pouliquen, O., Nicolas, M., 2020. Cohesion-controlled granular material. Phys Rev E 101, 32904. https://doi.org/10.1103/PhysRevE.101.032904

Garcia, X., Latham, J.-P., Xiang, J., Harrison, J.P., 2009. A clustered overlapping sphere algorithm to represent real particles in discrete element modelling. Géotechnique 59, 779–784. https://doi.org/10.1680/geot.8.T.037

Gravish, N., Franklin, S. V, Hu, D.L., Goldman, D.I., 2012. Entangled granular media. Phys Rev Lett 108, 1–4. https://doi.org/10.1103/PhysRevLett.108.208001

Jaeger, H.M., Nagel, S.R., Behringer, R.P., 1996. Granular solids, liquids, and gases. Rev. Mod. Phys. 68, 1259–1273. https://doi.org/10.1103/RevModPhys.68.1259

Karapiperis, K., Monfared, S., de Macedo, R.B., Richardson, S., Andrade, J.E., 2022. Stress transmission in entangled granular structures. Granul Matter 24, 1–15. https://doi.org/10.1007/s10035-022-01252-4

Karuriya, A.N., Barthelat, F., 2024. Plastic deformations and strain hardening in fully dense granular crystals. J Mech Phys Solids 186, 105597. https://doi.org/https://doi.org/10.1016/j.jmps.2024.105597

Karuriya, A.N., Barthelat, F., 2023. Granular crystals as strong and fully dense architectured materials. Proc Natl Acad Sci U S A 120. https://doi.org/10.1073/pnas.2215508120

Kruggel-Emden, H., Rickelt, S., Wirtz, S., Scherer, V., 2008. A study on the validity of the multi-sphere Discrete Element Method. Powder Technol 188, 153–165. https://doi.org/https://doi.org/10.1016/j.powtec.2008.04.037

Leonardo E. Silbert Deniz Ertaş, G.S.G.T.C.H.D.L., Plimpton, S.J., 2001. Granular flow down an inclined plane: Bagnold scaling and rheology. Phys. Rev. E 64, 51302. https://doi.org/10.1103/PhysRevE.64.051302

Li, C.-Q., Xu, W.-J., Meng, Q.-S., 2015. Multi-sphere approximation of real particles for DEM simulation based on a modified greedy heuristic algorithm. Powder Technol 286, 478–487. https://doi.org/https://doi.org/10.1016/j.powtec.2015.08.026

Li, C.X., Zou, R.P., Pinson, D., Yu, A.B., Zhou, Z.Y., 2020. An experimental study of packing of ellipsoids under vibrations. Powder Technol 361, 45–51. https://doi.org/10.1016/j.powtec.2019.10.115

Marschall, T.A., Franklin, S. V, Teitel, S., 2015. Compression- and shear-driven jamming of U-shaped particles in two dimensions. Granul Matter 17, 121–133. https://doi.org/10.1007/s10035-014-0540-2

Mlot, N.J., Tovey, C.A., Hu, D.L., 2011. Fire ants self-assemble into waterproof rafts to survive floods. Proc Natl Acad Sci U S A 108, 7669–7673. https://doi.org/10.1073/pnas.1016658108

Murphy, K.A., Reiser, N., Choksy, D., Singer, C.E., Jaeger, H.M., 2016. Freestanding loadbearing structures with Z-shaped particles. Granul Matter 18. https://doi.org/10.1007/s10035-015-0600-2

Nicolas, M., Duru, P., Pouliquen, O., 2000. Compaction of a granular material under cyclic shear. The European Physical Journal E 3, 309–314. https://doi.org/10.1007/s101890070001

Onoda, G.Y., Liniger, E.G., 1990. Random loose packings of uniform spheres and the dilatancy onset. Phys. Rev. Lett. 64, 2727–2730. https://doi.org/10.1103/PhysRevLett.64.2727

Papadopoulos, L., Porter, M.A., Daniels, K.E., Bassett, D.S., 2018. Network analysis of particles and grains. J Complex Netw 6, 485–565. https://doi.org/10.1093/COMNET/CNY005

Philipse, A.P., 1996. The Random Contact Equation and Its Implications for (Colloidal) Rods in Packings, Suspensions, and Anisotropic Powders. Langmuir 12, 5971. https://doi.org/10.1021/la960869o

Ritchie, R.O., 2011. The conflicts between strength and toughness. Nat Mater 10, 817–822. https://doi.org/10.1038/nmat3115

Rodriguez, R.N., 1977. A guide to the Burr type XII distributions. Biometrika 64, 129–134. https://doi.org/10.1093/biomet/64.1.129

Smith, B.L., Schäffer, T.E., Viani, M.B., Thompson, J.B., Frederick, N.A., Kindt, J.H., Belcher, A.M., Stucky, G.D., Morse, D.E., Hansma, P.K., 1999. Molecular mechanistic origin of the toughness of natural adhesives, fibres and composites. Nature 399, 761–763.

Thompson, A.P., Aktulga, H.M., Berger, R., Bolintineanu, D.S., Brown, W.M., Crozier, P.S., in 't Veld, P.J., Kohlmeyer, A., Moore, S.G., Nguyen, T.D., Shan, R., Stevens, M.J., Tranchida, J., Trott, C., Plimpton, S.J., 2022. LAMMPS - a flexible simulation tool for particle-based materials modeling at the atomic, meso, and continuum scales. Comput Phys Commun 271, 108171. https://doi.org/https://doi.org/10.1016/j.cpc.2021.108171

Tsai, J.-C., Gollub, J.P., 2004. Slowly sheared dense granular flows: Crystallization and nonunique final states. Phys. Rev. E 70, 31303. https://doi.org/10.1103/PhysRevE.70.031303

Wagner, R.J., Such, K., Hobbs, E., Vernerey, F.J., 2021. Treadmilling and dynamic protrusions in fire ant rafts. J R Soc Interface 18, 20210213. https://doi.org/10.1098/rsif.2021.0213

Wang, Y., Li, L., Hofmann, D., Andrade, J.E., Daraio, C., 2021. Structured fabrics with tunable mechanical properties. Nature 596, 238–243. https://doi.org/10.1038/s41586-021-03698-7

Weiner, N., Bhosale, Y., Gazzola, M., King, H., 2020. Mechanics of randomly packed filaments—The "bird nest" as meta-material. J Appl Phys 127, 50902. https://doi.org/10.1063/1.5132809

Zhang, H.P., Makse, H.A., 2005. Jamming transition in emulsions and granular materials. Phys. Rev. E 72, 11301. https://doi.org/10.1103/PhysRevE.72.011301